\documentclass[12pt,a4paper]{article}
\usepackage{a4wide}
\usepackage{latexsym}
\usepackage{epsf}
\usepackage{amssymb}
\usepackage[english]{babel}
\usepackage{amsfonts}
\usepackage{amsmath}
\usepackage{fancyhdr}
\usepackage{floatflt,graphicx}
\usepackage{psfrag}
\usepackage{cite}
\begin{document}
\def\be{\begin{equation}}
\def\ee{\end{equation}}
\def\bea{\begin{eqnarray}}
\def\eea{\end{eqnarray}}
\def\pd{\partial}
\def\a{\alpha}
\def\b{\beta}
\def\g{\gamma}
\def\d{\delta}
\def\m{\mu}
\def\n{\nu}
\newcommand{\fsl}{{\hspace{-7pt}\slash}}
\newcommand{\gsl}{{\hspace{-5pt}\slash}}
\newcommand{\dslash}{\pd\fsl}
\newcommand{\pslash}{p\gsl }
\newcommand{\Aslash}{A\gsl }
\newcommand{\kslash}{k\gsl }
\newcommand{\Lslash}{\Lambda\gsl }
\newcommand{\kuslash}{k_1\gsl }
\newcommand{\kdslash}{k_2\gsl }
\newcommand{\Dslash}{D\fsl}
\def \h{\mathcal{H}}
\def \hh{\mathcal{G}}
\def\bi{\begin{itemize}}
\def\ei{\end{itemize}}
\def\t{\tau}
\def\p{\pi}
\def\th{\theta}
\def\l{\lambda}
\def\O{\Omega}
\def\r{\rho}
\def\s{\sigma}
\def\e{\epsilon}
  \def\scri{\mathcal{J}}
\def\cM{\mathcal{M}}
\def\tcM{\tilde{\mathcal{M}}}
\def\RR{\mathbb{R}}
\def\tr{\operatorname{tr}}
\def\str{\operatorname{str}}
\begin{flushright}
IFT-UAM/CSIC-07-05\\
hep-th/0702184\\
\end{flushright}
\vspace{1cm}
\begin{center}
{\bf\Large  Unimodular cosmology and the weight of energy}\\
\vspace{.5cm}
{\bf Enrique \'Alvarez and Ant\'on F. Faedo }
\vspace{.3cm}
\vskip 0.4cm  
 
{\it  Instituto de F\'{\i}sica Te\'orica UAM/CSIC, C-XVI,\\
and \\ Departamento de F\'{\i}sica Te\'orica, C-XI,\\
  Universidad Aut\'onoma de Madrid 
  E-28049-Madrid, Spain }
\vskip 0.2cm
\vskip 1cm
\begin{abstract}
Some models are presented in which the strength of the gravitational coupling of the 
potential energy relative
to the same coupling for the kinetic energy is, in a precise sense, adjustable. 
The gauge symmetry of these 
models consists
of those coordinate changes with unit  jacobian.
\end{abstract}
\end{center}
\begin{quote}
  
\end{quote}
\newpage
\setcounter{page}{1}
\setcounter{footnote}{0}
\tableofcontents
\newpage
\section{Introduction}
There are several aspects of the cosmological constant
problem.  One of them is  the dynamical nature of  dark energy, often parameterized by
its equation of state $p=w \rho$. This shall be refered to as the {\em inverse cosmological 
constant problem}, in the sense that there are data to be explained by a theoretical
construct. A pure cosmological constant 
corresponds to $w=-1$. This aspect
is mainly cosmological, and progress on it can hopefully be made by refining the 
observational data, which nowadays seem to strongly favor  $\Omega_{\Lambda}\sim 0.7$ 
corresponding to a mass scale
\be
M_{DE}\sim 10^{- 12} GeV
\ee
and $w\sim -1$
\cite{Spergel}. 
\par
Even if a dynamical component such as quintessence is discovered that
explains fully the acceleration of the Universe, other problems remain.
For example, it  would
still not  be understood why the vacuum energy due to spontaneous symmetry 
breaking in quantum field theory does
 not produce a much larger cosmological constant. This shall be designated as the 
{\em direct cosmological constant problem}, in the sense that it refers to an unfulfilled 
theoretical prediction.
\par
 A drastic possibility is that the way spontaneous 
symmetry breaking occurs in quantum physics is not understood at all. 
It has been suggested, in particular by R.L. Jaffe \cite{Jaffe}, that even the experimental 
confirmation of the Casimir
 energy \cite{Lamoreaux} should not be considered as a proof of the existence 
of (zero point) vacuum energy in 
quantum field
theory. This seems to be an extreme viewpoint. A clear experimental signal of vacuum energy 
as well of the strength of its gravitational coupling would be most welcome. 
\par
As a matter of fact, even allegedly
uncontroversial breakings contribute to the problem. For instance, chiral symmetry 
breaking due to the quark condensate
yields a contribution of the order of $\Lambda_{QCD}$, which is eleven orders of magnitude 
\footnote{ Those figures get multiplied by a factor four if the energy density is the 
quantity to be compared.}
above the purported range of observed values
for the cosmological constant.
\par
Turning again to the general aspects of the  direct problem, the status is as follows.
 In order to not strongly disagree with observations,
there are essentially two possibilities: either there is a cancellation in such a way that the
low energy vacuum energy is very small (in natural units) or else there is a modification of
gravity such that vacuum energy does not gravitate (or gravitates less than ordinary matter).
This viewpoint has been advocated, for example, in \cite{Arkani-Hamed}, but the theories 
proposed there are badly non-local. Still (as has also been hinted at by those authors) there
is an intriguing relationship between their approach and unimodular theories in general
 which has not 
been fully elucidated yet.
\par
The purpose of the present work is to explore in some detail a  setting in which  the relative 
weight of the vacuum energy with respect to the kinetic energy can be tuned at will, and indeed
 in an extreme case the
vacuum energy does not weigh at all. This will be done in a local theory that is a 
minor modification of General Relativity (GR), in the sense that the gauge group is not
the full set of general coordinate transformations (GC) (which will be interpreted
in the active sense as diffeomorphisns  spanning the group Diff(M)), but rather those that 
enjoy unit jacobian. These particular transformations have been called {\em unimodular 
transformations}, and in a preceding paper the name 
transverse diffeomorphisms (TDiffs) spanning  a subgroup TDiff(M) has been used. 
Please refer to  \cite{Alvarezz} where other
 relevant references can also be found.
To be specific, transverse diffeomorphisms (TDiffs) in a spacetime manifold whose points 
are described in a 
particular coordinate chart by $x^\m(P)$, $\m=0,1,\ldots,n-1$ are those diffeomorphisms  
\be
x^\m\rightarrow y^\m
\ee
such that the determinant of the jacobian matrix equals unity:
\be
D(y,x)\equiv\det\,\left(\frac{\pd y^\m}{\pd x^{\n}}\right)=1
\ee
In the linearized approximation, the Diff is expressed through a vector field;
\be
y^\m=x^\m+\xi^\m(x)
\ee
and the above mentioned condition is equivalent to
\be
\pd_\a \xi^\a=0
\ee
This is the reason of the qualifier {\em  transverse} applied to them. 
Usually {\em tensor densities of weight $w$} are defined in such a way that they get an 
extra factor of the jacobian to the power $w$ in the tensorial transformation law.
For example, a scalar density transforms as a one-dimensional representation of Diff(M),
namely:
\be
\phi_w^{\prime}(y)=\left(D(y,x)\right)^{w}\phi(x)
\ee
Whenever there is a metric (in the usual sense, a rank two tensor and not a tensor density),
the determinant $g\equiv \det\, g_{\m\n}$ behaves as a
scalar density of weight $w=-2$.
\par
This means that as long as we assume that TDiff is the basic symmetry of Nature, we do not
distinguish tensor densities among themselves. In particular, given a certain
scalar field, $\phi(x)$, all dressed fields $f(g) \phi(x)$ also behave as scalars under TDiff.
\par
Vector fields inducing TDiffs can then be represented as
\be
\xi^\a=\e^{\a\m_2\ldots\m_n}\pd_{\m_2}\Omega_{\m_3\ldots \m_n}=\e^{\a\m_2\ldots\m_n}
\nabla_{\m_2}\Omega_{\m_3\ldots \m_n}
\ee
where $\e^{\m_1\m_2\ldots\m_n}$ is the contravariant Levi-Civita tensor, and 
$\Omega_{\m_3\ldots \m_n}$ is completely antisymmetric i.e., they are the components of a
 $(n-3)$-form.
\section{Some simple models }
We shall study a particular class of theories which enjoy TDiff (as opposed to full Diff)
invariance. As explained before, we can dress the gravitational and the matter sectors with 
different functions of the determinant of the metric, namely:
\be
S\equiv \int d^n x \left(-\frac{1}{2 \kappa^2}f(g) R + f_m(g)L_m(g_{\m\n},\phi^{(w_i)}_i,g)
\right)
\ee
In this formula, $\kappa^2\equiv 8\pi G_n$, $G_n$ being the n-dimensional Newton's constant. 
$\phi_i$ ($i=1\ldots N$) represent matter fields, which enjoy arbitrary weights $w_i$ on top of
their own transformation properties under the group Diff(M). The determinant of the 
metric, $g$, is counted
as a matter field with $w=-2$ for those purposes.
\par
The simplest instance posits $L_m$ as a full scalar, which will be 
assumed by simplicity not to depend on derivatives of the metric (so that all  matter fields are
minimally coupled). More general terms involving $\pd_\m g$ are not included for the time being 
since they introduce additional unnecessary complications. That is, the Lagrangian to be 
considered in the present paper is:
\be
S\equiv \int d^n x \left(-\frac{1}{2 \kappa^2}f(g) R + f_m(g)L_m(g_{\m\n},\phi_i)\right)
\ee
(with all weights $w_i=0$).
\par
It should be remarked from the start than this action principle is non fully covariant; if it is
 assumed valid in a certain reference systen (RS), in general coordinates reads:
\be\label{groupav}
S=\int d^n x \frac{1}{C(x)}\left(-\frac{1}{2 \kappa^2}f(g(x) C(x)^{2}) R(x) + 
f_m(g(x) C(x)^{2})
L_m(g_{\m\n}(x),\phi_i(x))
\right)
\ee
where $C(x)\equiv D(x,\bar{x})$ is the determinant of the jacobian matrix with respect
to some privileged original coordinates denoted here by $\bar{x}$. 
The field $C(x)$ can also be viewed as the determinant of the four vector fields 
$A^\m_{(\bar{\a})}$, constructed from
the derivatives of the coordinate functions with respect to the fiducial functions 
$\bar{x}^\m(x)$. As long as the field $C(x)$ is  kept
in the lagragian, formal Diff invariance is ensured by the chain property
\be
C(y)\equiv det\,\frac{\pd y}{\pd \bar{x}}=det\,\frac{\pd y}{\pd x} 
det\,\frac{\pd x}{\pd \bar{x}}=D(y,x)C(x)
\ee
i.e., the composite field
\be
g(x)C(x)^2
\ee
is a gauge invariant construct under the gauge group Diff(M).
This type of field is sometimes \cite{Gates} called a {\em compensator field}. A notorious
exaple is the Stueckelberg field which renders gauge invariant massive electrodynamics.
The original theory can always be recovered in the analogous of the {\em unitary} gauge
\be
C(x)=1
\ee 
\par
\subsection{The source of gravity}
There are several energy momentum tensors of interest in this case. Actually, they are in 
general not true tensors under GC; only densities.
The true energy-momentum tensor (that is, the source of the gravitational equations), is
\be
T_{\m\n}\equiv\frac{\d S_m}{\d g^{\m\n}}\equiv \frac{\d}{\d g^{\m\n}}\int d^n x f_m L_m
\ee
In order to study its conservation law, let us perform a TDiff (cf. \cite{Alvarezz}),
under which
\be
\d g_{\a\b}=\pounds(\xi)g_{\a\b}\equiv \xi^{\rho}\pd_\rho g_{\a\b}+g_{\a\rho}\pd_\b \xi^\rho 
+ g_{\rho\b}\pd_\a \xi^\rho
\ee
The use of covariant derivatives is best avoided for the time being. The fact that the
quantities considered are not tensors under Diff has already been mentioned, and this can 
obscure the reasoning. Performing a TDiff on the matter action 
\bea
0&=&\d_{T-diff}S_m=\int d^n x \left(\e^{\rho\m_2\ldots\m_n}\pd_{\m_2}\Omega_{\m_3
\ldots \m_n}\pd_{\rho} g_{\a\b}+
g_{\a\rho}\pd_\b\left(\e^{\rho\m_2\ldots\m_n}\pd_{\m_2}\Omega_{\m_3
\ldots \m_n}\right)\right. \nonumber\\
&&\left.+g_{\b\rho}\pd_\a\left(\e^{\rho\m_2\ldots\m_n}\pd_{\m_2}\Omega_{\m_3
\ldots \m_n}\right)\right) T^{\a\b}
\eea
Taking into account that $\e^{\m_1\ldots\m_n}$ is independent of the metric, and denoting
\be
\omega^{\m\n}\equiv \e^{\m\n\m_3\ldots\m_n}\Omega_{\m_3\ldots\m_n}
\ee
the aforementioned condition is equivalent to:
\be
0=\int d^n x\,\omega^{\m\n}\left(-\pd_\m g_{\a\b}\pd_\n 
T^{\a\b}+2\pd_\n \pd_\l T_\m\,^\l\right) 
\ee
This means that
\be\label{cons}
\pd_\m g_{\a\b}\pd_\n T^{\a\b}-\pd_\n g_{\a\b}\pd_\m T^{\a\b}=
2\left(\pd_\n \pd_\l T_\m\,^\l-\pd_\m \pd_\l T_\n\,^\l\right)
\ee
which does imply 
\be\label{fi}
\pd_\l T_\m\,^\l-\frac{1}{2}\pd_\m g_{\a\b}T^{\a\b}=\pd_\m \Phi
\ee
where $\Phi$ is an arbitrary function. 
Using the well-known formula (valid for any symmetric tensor)\footnote{
This formula is attributed by Eisenhart in \cite{Eisenhart} to none other than Einstein himself
(1916).}
\be
\nabla_\n S_\m\,^\n=\frac{1}{\sqrt{|g|}}\pd_\n\left(\sqrt{|g|}S_\m\,^\n\right)-\frac{1}{2}
\pd_\m g_{\a\b}S^{\a\b}
\ee
this can be rewritten as
\be\label{bianchi}
\nabla_\n\left(\frac{T_\m\,^\n}{\sqrt{|g|}}\right)=\frac{1}{\sqrt{|g|}}\,\,\pd_\m\Phi
\ee
in the understanding that the covariant derivative is to be taken as if  $T_{\a\b}$ were a 
true tensor. Note in passing that (\ref{cons}) also implies
\be
\pd_\l T_\m\,^\l+\frac{1}{2}g_{\a\b}\pd_\m T^{\a\b}=\pd_\m \Phi^{\prime}
\ee
And the difference between both arbitrary functions is just the trace of the 
energy-momentum tensor
\be
\Phi^{\prime}-\Phi=g_{\a\b}T^{\a\b}=T
\ee
On the other hand, it is clear from its definition that
\be\label{tmunu}
T_{\m\n}=f_m\frac{\d L_m}{\d g^{\m\n}}
- g f^{\prime}_m L_m g_{\m\n}
\ee
We have used the abbreviation $\frac{\d L_m}{\d g^{\m\n}}$ instead of the most accurate 
$\frac{\d \int d^n x L_m}{\d g^{\m\n}}$. It is interesting to study the nature of this tensor
since as we have seen it is not possible to deduce its covariant consevation using only 
invariance under TDiffs.
\par
\subsection{Energy-momentum tensors}
The label {\em energy-momentum tensor} for the above construct (\ref{tmunu}) can indeed be 
questioned for
very good reasons. It is a metric (Rosenfeld) tensor which is not conserved, and consequently,
it does not reduce in flat space to the canonical one, or to its equivalent Belinfante form
(cf. \cite{Ortin} for a lucid discussion of the standard situation). That is, the tensor
(\ref{tmunu}) does not convey the Noether current corresponding to translation invariance.
In order to illustrate this, let us consider the simplest example, namely a real scalar field
without coupling to the determinant of the metric, i.e., $f_m(g)=1$
\be
S_m\equiv\int d^n x L_m =\int d^n x \frac{1}{2} g^{\m\n}\pd_\m\phi\pd_\n\phi
\ee 
The energy-momentum tensor as defined before is
\be
T_{\m\n}=\frac{1}{2}\pd_\m\phi\pd_\n\phi
\ee
Using the equation of motion (EM) of the scalar
\be
\frac{\d S_m}{\d \phi}\equiv\pd_\m\left(g^{\m\n}\pd_\n\phi\right)=0
\ee
it can be shown that
\be\label{conser}
\sqrt{|g|}\nabla_\n\left(\frac{T_\m\,^\n}{\sqrt{|g|}}\right)=\frac{1}{2}\nabla_\m L_m
\ee
conveying that fact that this energy momentum is not covariantly conserved, and
thus it cannot act as a consistent source of Einstein's equations. 
What is worse, $T_{\m\n}$ does not reduce in flat
space to the canonical one
\be
T_{\m\n}^{can}=\pd_\m\phi\pd_\n\phi-\frac{1}{2}L_m \eta_{\m\n}
\ee
which is well known to be conserved.
This does not happen of course with the usual covariant lagrangian
\be
S_{cov}=\int d^n x \sqrt{|g|}\frac{1}{2}g^{\m\n}\pd_\m\phi\pd_\n\phi
\ee
whose energy-momentum tensor
\be
T_{\m\n}^{GR}\equiv\frac{2}{\sqrt{|g|}}\left(\frac{1}{2}\sqrt{|g|}\pd_\m\phi\pd_\n\phi-
\frac{1}{4}\sqrt{|g|}g_{\m\n}g^{\a\b}\pd_\a\phi\pd_\b\phi\right)
\ee
is both covariantly conserved thanks to the new EM and reduces to the canonical one in flat space.
\subsection{The GR template}
It is instructive to check all this setup applying it to the case of General Relativity.
On general grounds Diff(M) invariance forces $f=f_m=\sqrt{|g|}$. The corresponding
energy momentum tensor is then
\be
T^{Diff}_{\m\n}=\sqrt{|g|}\left(\frac{\d L_m}{\d g^{\m\n}}
- \frac{1}{2} L_m g_{\m\n}\right)
\ee
As an exercise we will derive the well-known conservation of the GR energy-momentum tensor
with our present techniques. Performing a Diff on the matter action and demanding it to be stationary
\be
0=\d_{Diff} S=\int d^n x\left(\xi^\rho \pd_\rho g_{\a\b}+g_{\a\rho}\pd_\b\xi^\rho+
g_{\rho\b}\pd_\a\xi^\rho\right)T_{Diff}^{\a\b}
\ee
Conveying the fact that
\be
0=\pd_\rho g_{\a\b}T_{(Diff)}^{\a\b}-\pd_\b T^{(Diff)}_\rho\,^\b-\pd_\a T_{(Diff)}^\a\,_\rho=
-2\sqrt{|g|}\nabla_\a\left(
\frac{T^{(Diff)}_\rho\,^\a}{\sqrt{|g|}}\right)
\ee
Please note that precisely for this reason the energy-momentum tensor is usually 
defined without the $\sqrt{|g|}$ factor (and with a conventional factor of 2 as well).
\be
T^{GR}_{\a\b}\equiv\frac{2}{\sqrt{|g|}}T_{\a\b}^{Diff}
\ee
\section{The gravitational equations of motion}
Once we have discussed in detail the source of gravity in these models, let us turn our 
attention to the complete equation. Including in the Einstein--Hilbert sector an arbitrary weight $f(g)$,
the gravitational equation of motion reads:
\be\label{filipina}
\frac{\d S}{\d g^{\m\n}}=-\frac{1}{2\kappa^2}\left(fR_{\m\n}-g f^{\prime} R g_{\m\n}-\sqrt{|g|}\left(
\nabla_{(\m}\nabla_{\n)}-\nabla^2 g_{\m\n}\right)\frac{f}{\sqrt{|g|}}\right)+  T_{\m\n}=0
\ee
It is worth stressing that the EM (\ref{filipina}) is not a tensor equation with respect to GC.
This means that it must be solved in one reference system (RS), and the results in 
another RS are not
the same if the transformation from one to the other is not a TDiff.
This poses interesting problems of principle, many of them already discussed some time ago
(cf. for example, the discussion of harmonic coordinates in \cite{Fock}).\footnote{It 
is worth noticing that when
\be
f=f_m=|g|^{1/n}
\ee
the EM are somewhat similar to Einstein's 1919 traceless equations (cf.\cite{Alvarezz}), namely
\be\label{tracefree}
R_{\m\n}-\frac{1}{n}R g_{\m\n}=|g|^{\frac{n-2}{2n}}\left(
\nabla_{(\m}\nabla_{\n)}-\nabla^2 g_{\m\n}\right)|g|^{\frac{2-n}{2n}}+
2\kappa^2 |g|^{-1/n}T_{\m\n}
\ee
The second member is not automatically tracefree; consistency demands that
\be
\left(1-n\right)\sqrt{|g|}\nabla^2 |g|^{-\frac{n-2}{2}}+2\kappa^2 T=0
\ee
}
\par
 
In the absence of matter (implying $T_{\m\n}=0$ since we have not considered
a cosmological constant term), the trace reads
\be
\left(1- n \frac{g f^{\prime}}{f} \right)R=(1-n)\frac{\sqrt{|g|}}{f}\nabla^2 
\frac{f}{\sqrt{|g|}}
\ee
In the simplest case that $f=\sqrt{|g|}$ the equation is tensorial in character
and coincides with the GR case. This automatically implies Ricci flatness, supposing 
$n\ne 2$. Concerning the general case, $f\neq \sqrt{|g|}$, in the absence of sources from the 
preceding trace equation we are not able to deduce that the scalar curvature vanishes.  
This is generically incompatible with the well-known solar system test of GR and
is the main reason why we consider $f=\sqrt{|g|}$ in the following.
The equation with sources is then
\be\label{sources}
G_{\m\n}=\frac{2\kappa^2}{\sqrt{|g|}} T_{\m\n}
\ee
\par
A small paradox can be now disposed of. Let us assume, as
seem obvious, that $G_{\m\n}$ is really a tensor, the Einstein tensor. Then it must obey 
Bianchi identities, which ensure that
\be
\nabla_\a G^\a{}_\m=0
\ee
so from the previous equation (\ref{sources})
\footnote{When the true physical
invariance is restricted TDiff, the same argument we used to arrive to (\ref{bianchi}) leads to
\be
\nabla_\l \left(\frac{E_\m\,^\l}{\sqrt{|g|}}\right) =\frac{1}{\sqrt{|g|}}\,\,
\pd_\m \Sigma
\ee
where
\be
E_\m\,^\l\equiv f(g) R_\m\,^\l-g f^{\prime} R \d^\l_\m-\sqrt{|g|}\left(
\nabla^\l\nabla_\m-\nabla^2 \d^\l_\m\right)\frac{f}{\sqrt{|g|}}
\ee
is such that, when $f =\sqrt{|g|}$, it reduces to the Einstein tensor multiplied by the
square root of the determinant of the metric:
\be
E_\m\,^\l=\sqrt{|g|}G_\m\,^\l
\ee
and the gravitational equations of motion simply demand that
\be
\pd_\m\left(\Sigma-2\kappa^2\Phi\right)=0
\ee
It is nevertheless true that when $f=\sqrt{|g|}$ there is an enhanced Diff symmetry in the
pure gravitational sector, so that it is to be expected that Bianchi identities remain valid.
}
\be\label{filipinaa}
\nabla_\a\left(\frac{T^\a{}_\m}{\sqrt{|g|} }\right)=0
\ee
precisely the same integrability condition that appears in GR, where Diff invariance
combined with the EM of matter imply covariant conservation of the energy-momentum tensor
independently of Bianchi identities.
\par
 From the TDiff viewpoint this is an extra condition that should be added for consistency
 but one that looks a bit mysterious.
\par
\subsection{The r\^ole of the compensators.}
In order to clarify the situation, let us consider the formally invariant theory
defined with the compensator field. Introducing the compensator $C(x)$ changes
the matter Lagrangian in the simplest case of a scalar to
\be
S_m=\int \frac{d^n x}{C(x)}\,\,f_m\left(g(x)C(x)^2\right)\,\, L_m
\ee
and the EM for the compensator is simply
\be\label{constr}
\frac{\d S_m}{\d C(x)}=-\frac{1}{C^2}\,\,f_m\,\,L_m+\frac{1}{C}\,\,\frac{\pd f_m}{\pd C}\,\,L_m=0
\ee
being the only solutions either $L_m=0$ or $f_m\sim C$. This last solution implies
$f_m\sim\sqrt{|g|}$, as one should expect because it corresponds to Diff invariance and the
compensator is just looking for it. If we do not want to impose Diff invariance from the
very beginning, we are forced to impose the constraint
\be
L_m=0
\ee
The meaning of this is that in this sector, the TDiff energy-momentum tensor is a consistent
source of Einstein's equations: the constraint ensures the integrability condition
(\ref{filipinaa}) to hold, although it is somewhat stricter than necessary
 (see (\ref{conser})). In any case, from this more elegant point of view,
conservation of the energy-momentum tensor in the sense of (\ref{filipinaa}) is just a consequence 
of compensator dynamics.
\par
Similar results are obtained when the model is generalized by allowing matter 
fields $\phi^{(w)}$ to become scalar densities of weight $w$ and considering Diff invariant Lagrangians
\be
S_m=\int d^nx\frac{1}{C(x)}\,\,f_m\left(g(x)C(x)^2\right)\,\,L_m(\phi^{(w)}C(x)^{-w})
\ee
The EM for the compensator now takes the form
\be
\frac{\d S_m}{\d C(x)}=-\frac{1}{C^2}\,\,f_m\,\,L_m+\frac{1}{C}\,\,\frac{\pd f_m}{\pd C}\,\,L_m+\frac{1}{C}\,\,f_m\,\,\frac{\d L_m}{\d C}=0
\ee
But given the functional dependence of $L_m$ on the combination $\phi^{(w)}C(x)^{-w}$
it is easy to convince oneself that 
\be
\frac{\d L_m}{\d C}\propto \frac{\d L_m}{\d \phi^{(w)}}
\ee
Therefore, (\ref{constr}) is recovered under the assumption that the scalar field verifies its EM.
\section{TDiff Friedmann models }
\subsection{Weight one}
Let us study in detail the extreme case in which the matter Lagrangian does not
couple to the determinant of the metric. This means $f_m=1$ in our previous notation.
\par
Lat us stress that we do not expect these models to be realistic. In practice $f_m\sim f_g$. Our aim in this paragraph
is a sort of {\em existence proof}; i.e., to show that it is possible to build theoretically consistent models
in the unimodular framework in which the potential energy does not weigh at all, and consequently, 
models that  solve the direct cosmological constant problem. 
\par
The simplest example is again a minimally coupled scalar field:
\be\label{scalar}
L_m=\frac{1}{2}g^{\m\n}\pd_\m\phi\pd_\n\phi-V(\phi)
\ee
where $V(\phi)$ is a polynomial representing the potential energy, and which does not
contribute to the gravitational EM (this is the main motivation for assuming $f_m=1$). 
This point is so important that it is worth emphasizing: not all energy is a source 
of the gravitational field in TDiff theories, but only the kinetic part of it.
\par
The preceding paragraph would be exactly true were it not for the compensators. Actually,
all energy interacts with the gravitational field through compensator exchange.
 This point shall be hopefully clarified in the sequel.
\par
In the final section of this paper we will introduce a framework which allows in principle 
an experimental test of this hypothesis and, in general, of the strength of the coupling
of the potential energy to the gravitational field.
\par
The energy-momentum tensor reads:
\be
T_{\m\n}=\frac{1}{2}\pd_\m\phi\pd_\n\phi
\ee
This means that it is not only the constant piece, but rather the full potential energy
density that does not generate gravitational field. The matter EM are:
\be
\pd_\m\left(g^{\m\n}\pd_\n \phi\right)+V^{\prime}(\phi)=0
\ee
Plus the EM for the compensator, which is again 
\be
L_m=0
\ee
This equation is the one that feeds back the potential energy into the gravitational equations,
although in a very unusual way.
\subsection{Friedmann}
In order to get an idea of the simplest cosmological consequences of the TDiff viewpoint,
 it would be convenient to assume a (spatially) flat Friedmann metric:
\be
ds^2 = dt^2-R(t)^2\d_{ij}dx^i dx^j\equiv dt^2-R(t)^2\left(dx^2+ dy^2 + dz^2\right)
\ee
This particular form of the metric is written however in the so called synchronous gauge
\bea
&&g_{00}=1\nonumber\\
&&g_{0i}=0
\eea
(defining what is known as Robertson-Walker\footnote{
The use of a specific set of Robertson-Walker coordinates is not as innocent 
as it seems, and physics 
does depend on this choice to a certain extent in the present framework.} coordinates)
which is not accessible in general using TDiffs only, since we have now less gauge
freedom than in General Relativity because of the transversality condition, 
or equivalently we have
already  partially fixed the gauge by choosing $C(x)=1$.
\par
The simplest form of a metric that can be reached with TDiff is
\be\label{metric}
ds^2 = a(t) dt^2-R(t)^2\d_{ij}dx^i dx^j\equiv a(t) dt^2-R(t)^2\left(dx^2+ dy^2 + dz^2\right)
\ee
The corresponding Einstein tensor is
\bea
G_{00}&=&3\frac{\dot{R}^2}{R^2}\equiv 3 H(t)^2\nonumber\\
G_{ij}&=&\frac{\dot{a}\dot{R}R-2a\ddot{R}R-a\dot{R}^2}{a^2}\,\,\d_{ij}
\eea
So the spacelike sector of Einstein's equations then imply that 
\be
\pd_i\phi=0
\ee
and the gravitational equations take the simple form (using the compensator EM)
\bea\label{graveq}
3 H^2&=&\frac{\kappa^2}{a^{1/2}R^3}\dot{\phi}^2=\frac{2 a^{1/2}\kappa^2}{R^3}V(\phi)\nonumber\\
3aH^2+2a\dot{H}-\dot{a}H&=&0
\eea
Since we have now an additional constraint imposed by the EM of the compensator, or in other language
by the conservation of the energy-momentum tensor, it is interesting to check the compatibility
of the whole system of equations. The usual counting of degrees of freedom in GR is as follows.
In four dimensions the metric has $10$ independent components, $4$ of which can be gauge fixed.
Then one has $6$ variables for $10$ Einstein's equations, but again using the Bianchi identities it is
certain that four of them are combination of the others, so finally one 
is left with
$6$ equations for $6$ variables.

When we consider a TDiff model such as the one above, the freedom to fix the gauge is 
smaller so that we
have to determine $7$ components of the metric. The Bianchi identities are nevertheless satisfied and
then there are $6$ Einstein's equations. The consistency is saved finally by the EM of the compensator
that provides the $7^{th}$ equation for the $7$ variables.
Let us prove this assertion in the particular model considered, i.e., the scalar Lagrangian (\ref{scalar})
with the metric (\ref{metric}). Conservation of the energy-momentum tensor forces the Lagrangian to 
be a constant, so differentiating (\ref{scalar}) with respect to time
\be
\frac{1}{2}\frac{d}{dt}\left(\frac{\dot{\phi}^2}{a}\right)-V^{\prime}(\phi)\dot{\phi}=0
\ee
Using the EM of the scalar to eliminate the potential, after some straightforward algebra
we get the condition (supposing $\dot{\phi}\ne 0$)
\be
\frac{\ddot{\phi}}{\dot{\phi}}=\frac{3}{4}\frac{\dot{a}}{a}
\ee
On the other hand, the first of the Einstein's equations in (\ref{graveq}) implies
\be
\frac{\ddot{\phi}}{\dot{\phi}}=\frac{\dot{H}}{H}+\frac{3}{2}\frac{\dot{R}}{R}+\frac{1}{4}\frac{\dot{a}}{a}
\ee 
Finally, equating these last two expressions allows us to derive the second Einstein's equation in (\ref{graveq}).
We conclude that the system of equations must be compatible and the three functions ($\phi(t)$, $a(t)$, $R(t)$)
can be determined. This same consistency is not found if we choose $a(t)\equiv 1$ from the beginning\footnote{The inconsistency is between conservation of the energy-momentum tensor, or compensator's EM, and the scalar EM.}, except
in the particular case in which the potential is a constant, and this
will be related with the impossibility to get exponential expansion, as we will see.

It is easy to find the explicit solution in the absence of potential. If the matter Lagrangian is
fixed to be a constant $L_m=L$ (the freedom to fix the constant is lost when the EM for the compensator is used
so that $L=0$), then the component of the metric $a=a_0$ is also constant and the scale
factor goes as 
\be
H(t)=\frac{H_0}{1+\frac{3}{2}H_0\,t}\hspace{1cm}\Longrightarrow\hspace{1cm} R(t)=R_0\left(1+\frac{3}{2}H_0\,t\right)^{\frac{2}{3}}
\ee
where $R_0^3=\frac{2a_0^{1/2}\kappa^2L}{3H_0^2}$. The time evolution of the scalar is linear and given by
\be
\phi(t)=\phi_0\pm \sqrt{2La_0}\,\,\, t
\ee
Once we choose to impose the constraint $L=0$ the only solution if the potential vanishes
is a constant field with null Hubble constant, but $a$ remains undetermined.

\subsection{The GR template}
These equations ought to be contrasted with the standard GR ones, where the synchronous gauge
$a=1$ is fully accessible. The symmetry of the situation enforces again $\pd_i\phi=0$ so that
for a general potential they read
\bea\label{grav}
3 H(t)^2&=&\kappa^2\left(\frac{1}{2}\dot{\phi}^2 + 
V(\phi)\right)\nonumber\\
2\dot{H}(t)+3 H(t)^2&=&-\kappa^2\left(\frac{1}{2}\dot{\phi}^2-V(\phi)
\right)
\eea
together with the GR equation of motion for the scalar field 
\be
\ddot{\phi}+ 3 H(t)\dot{\phi}+V^{\prime}(\phi)=0
\ee
\par
When the potential vanishes, the system is easy to solve giving a Hubble constant
\be
H(t)=\frac{H_0}{1+3H_0t}\hspace{1cm}\Longrightarrow\hspace{1cm} R(t)=R_0\left(1+3H_0\,t\right)^{\frac{1}{3}}
\ee
and a time dependence for the scalar field
\be
\phi(t)=\phi_0\pm\frac{\sqrt{6}}{3\kappa}\log \left(1+3H_0\,t\right)
\ee
\par
On the other hand, solutions that describe an exponentially expanding universe are 
very interesting phenomenologically. It is well known what is the origin of exponential expansion in GR: a positive 
constant energy density $V_0$ yields
\be
R(t)=R_0\, e^{\kappa \sqrt{\frac{V_0}{3}}t}
\ee

\subsection{No TDiff exponential expansion}
It is of interest to examine now the conditions under which there is exponential expansion 
in TDiff cosmology; that is, conditions for which TDiff behaves physically in a way similar 
to GR with a cosmological constant. When $\dot{H}=0$ but $H=H_0$ itself is nonvanishing
\be
R=R_0e^{H_0t}
\ee
and the second of the equations (\ref{graveq}) yields
\be
a=a_0 e^{3 H_0 t}
\ee
We can immediately solve for the scalar
\be
\phi(t)=\phi_0\pm\frac{4}{9\kappa}\left(3a_0^{\frac{1}{2}}R_0^3\right)^\frac{1}{2}\left(e^{\frac{9}{4}H_0t}-1\right)
\ee
which corresponds (using the compensator's EM) to a potential
\be
V(\phi)=\frac{3H_0^2R_0^3}{2\kappa^2a_0^{\frac{1}{2}}}\left(1\pm\frac{9\kappa}{4\left(3a_0^{\frac{1}{2}}R_0^3\right)^{\frac{1}{2}}}\left(\phi-\phi_0\right)\right)^{2/3}
\ee
This expansion, is, however not exponential with respect to comoving proper  time defined by
\be
dT=a^{1/2} dt
\ee
In order for the expansion to be exponential in $T$, it would have to obey
\be
\frac{dR}{a^{1/2} dt}= R H_0
\ee
This yields in the second equation of the set (\ref{graveq})
\be
3 a^2 H_0^2=0
\ee
This means that there is no truly (proper time) exponential expansion in this class 
of TDiff cosmologies, so that the direct cosmological constant problem appears in a new light: not only a
constant term in the potential does not gravitate, but there is no way to get the gravitational field which
is produced by such a term in GR. 

\subsection{Adjustable coupling gravity/potential energy.}
Let us consider the general case in which the matter Lagrangian (\ref{scalar}) is coupled to the
determinant of the metric to an arbitrary power, i.e., the action takes the form
\be
S_m=\int d^nx \,\,g^b \left(\frac{1}{2}g^{\m\n}\pd_\m\phi\pd_\n\phi-V(\phi)\right)
\ee
This framework allows to search for solutions that depend on the strength of the coupling 
between matter and the determinant of the metric parametrized in $b$ and at the end hopefully 
measurable physical consequences.

Conservation of the energy-momentum tensor in the sense of (\ref{filipinaa}) forces the 
Lagrangian to verify the integrability condition
\be\label{integra}
g^{b-\frac{1}{2}}\left(\frac{1}{2}-b\right)\pd_\m L_m=b\,\,L_m\,\,\pd_\m g^{b-\frac{1}{2}}
\ee
where we have used the scalar equation of motion
\be
\pd_\m\left(g^bg^{\m\n}\pd_\n\phi\right)+g^bV^{\prime}=0
\ee
The previous equation is less restringent than the one coming from the compensator EM,
 as we have already mentioned, but is also less motivated. Note that it is identically verified for the GR case which corresponds
to $b=\frac{1}{2}$. The condition of a constant Lagrangian corresponding to $b=0$, i.e., 
the model studied before, is also reproduced. Using these last two equations together with
timelike sector of Einstein's equations
\be
3H^2=2\kappa^2a^{b-\frac{1}{2}}R^{6\left(b-\frac{1}{2}\right)}a\left(\left(1-b\right)\frac{\dot{\phi}^2}{2a}+bV\right)
\ee
one can reproduce in a similar way as before the spacelike Einstein's equation 
\be
\dot{a}H-3aH^2-2a\dot{H}=2b\kappa^2a^{b-\frac{1}{2}}R^{6\left(b-\frac{1}{2}\right)}a^2\left(\frac{\dot{\phi}^2}{2a}-V\right)
\ee
under the assumption $b\ne1,\frac{1}{2}$. The GR limit has been studied in the previous 
subsection, and the
case $b=1$ is somehow special. Apart of these subtleties, the consistency of 
the system of equations
is ensured. In fact, in the absence of potential it is easy to find the solution
as a function of $b$
\be
H(t)=\frac{H_0}{1+\frac{3\left(b-\frac{1}{2}\right)}{\left(b-1\right)}H_0\,t}\hspace{1cm}
\Longrightarrow\hspace{1cm} R(t)=R_0\left(1+\frac{3\left(b-\frac{1}{2}\right)}{(b-1)}H_0\, t\right)^{\frac{(b-1)}{3\left(b-\frac{1}{2}\right)}}
\ee
Moreover, the temporal component of the metric is no longer a constant but
\be
a(t)=a_0\left(1+\frac{3\left(b-\frac{1}{2}\right)}{(b-1)}H_0\, t\right)^{\frac{2b}{(\frac{1}{2}-b)}}
\ee
and the scalar goes linearly in time independently of $b$
\be
\phi(t)=\phi_0\pm\sqrt{\frac{3}{\kappa^2(1-b)}}H_0\,\,a_0^{\frac{1}{2}(\frac{1}{2}-b)}R_0^{3(\frac{1}{2}-b)}\,\,t
\ee 

\section{Conclusions}
A family of models has been studied with slightly smaller  gauge symmetry
than the full set of general coordinate transformations  of General Relativity. 
Namely, they enjoy symmetry under unimodular  transformations , that is, 
diffeomorphisms with unit determinant of the jacobian matrix. They generate a group, that
is called TDiff(M).
\par
In some simplified TDiff cosmological models it has been found that  exponential 
expanding solutions
are inconsistent. This seems to alleviate the direct problem of the cosmological constant.
\par
The main reason however why those solutions are physically interesting is that they allow to 
tune, in a precise sense, the relative weight of the kinetic and potential energy. There are
models in particular, in which the potential energy (and {\em a fortiori} the vacuum 
energy) does not couple to the gravitational field, so that it appears that it does
not weight at all.\footnote{ Modulo some subtleties discussed in the main text, namely 
compensator exchange.}
This provides a framework to test the gravity/potential energy coupling, which
violates the {\em equivalence principle} inasmuch as it is different from the gravity/
kinetic energy coupling (cf. Damour's contribution in \cite{Damour}). Experiments 
are difficult 
but perhaps not impossible (weighing the Casimir energy?).
\par
The  vacuum gravitational equations  for the TDiff models studied in this paper are exactly 
the same as the GR ones, so 
that all solar systems
tests are also fulfilled. There might be some subtle points with the
derivation of the binary pulsar tests \cite{Weisberg} worth a detailed study.
 \par
Let us finally discuss if this framework could alter the results of an E\"otv\"os type experiment.
If a WKB expansion is performed \cite{Alvarez:2001qj} in the EM for the scalar field
\be
\phi=e^{\frac{i}{\e}\sum\e^n \phi_n}
\ee
then, to dominant order ($1/\e^2$), and defining $k_\m\equiv \pd_\m \phi_0$, 
the mass shell condition
\be
k^2= m^2
\ee
is recovered, as well as the geodesic equation in the form
\be
k^\a\nabla_\a k_\b=0
\ee
All this is quite similar to the GR template. The subtle differences in the coupling 
to the gravitational field in the scalar EM do not appear to this order
in the WKB expansion. The physical meaning of this result is that unimodular cosmology
predicts exactly the same results as GR for free falling of test bodies (to which
refer current experiments on the equivalence principle).
\par
Nevertheless, it has been claimed in \cite{Carlip} that the huge body of data on the  E\"otv\"os  experiment for different substances puts
constraints (barring accidental cancellations) on the possible violations of the equivalence principle on the potential energy 
(less than $10^{-10}$) and the kinetic energy separately (less than  $10^{-7}$). This then puts corresponding constraints on the ratio
$\frac{f_m}{f_g}$. Hypothetical positive results could be interpreted in the framework of our formalism.

\section*{Acknowledgments}
We are indebted to Diego Blas, Jaume Garriga, Bel\'en Gavela Eduard Mass\'o and Enric Verdaguer 
for illuminating discussions.
 This work has been partially supported by the
European Commission (HPRN-CT-200-00148) and by  FPA2006-05423 (DGI del MCyT, Spain).
A.F.F. has been supported by a MEC grant, AP-2004-0921.


\begin{thebibliography}{99}
\bibitem{Alvarezz}
  E.~Alvarez,
  ``Can one tell Einstein's unimodular theory from Einstein's general
  relativity?,''
  JHEP {\bf 0503}, 002 (2005)
  [arXiv:hep-th/0501146].
  E.~Alvarez, D.~Blas, J.~Garriga and E.~Verdaguer,
  ``Transverse Fierz-Pauli symmetry,''
  Nucl.\ Phys.\ B {\bf 756} (2006) 148
  [arXiv:hep-th/0606019].
\bibitem{Alvarez:2001qj}
  E.~Alvarez and J.~Conde,
  ``Are the string and Einstein frames equivalent?,''
  Mod.\ Phys.\ Lett.\  A {\bf 17} (2002) 413
  [arXiv:gr-qc/0111031].

\bibitem{Arkani-Hamed}
  N.~Arkani-Hamed, S.~Dimopoulos, G.~Dvali and G.~Gabadadze,
  ``Non-local modification of gravity and the cosmological constant problem,''
  arXiv:hep-th/0209227.
\bibitem{Carlip}
  S.~Carlip,
  ``Kinetic Energy and the Equivalence Principle,''
  Am.\ J.\ Phys.\  {\bf 65} (1998) 409
  [arXiv:gr-qc/9909014].
  
\bibitem{Carroll}
  S.~M.~Carroll,
  ``Dark Energy and the Preposterous Universe,''
  arXiv:astro-ph/0107571.
\bibitem{Damour}
  W.~M.~Yao {\it et al.}  [Particle Data Group],
  J.\ Phys.\ G {\bf 33} (2006) 1.
\bibitem{Eisenhart}L.P. Eisenhart, {\em Riemannian Geometry},
Princeton University Press (1964).
\bibitem{Fock}
V. Fock,
``The Theory of Space Time and Gravitation'',
(Pergamon, 1959)
\bibitem{Gates}
  S.~J.~Gates, M.~T.~Grisaru, M.~Rocek and W.~Siegel,
  ``Superspace, or one thousand and one lessons in supersymmetry,''
  Front.\ Phys.\  {\bf 58}, 1 (1983)
  [arXiv:hep-th/0108200].
\bibitem{Jaffe}
  R.~L.~Jaffe,
``The Casimir effect and the quantum vacuum,''
  Phys.\ Rev.\ D {\bf 72} (2005) 021301
  [arXiv:hep-th/0503158].
\bibitem{Lamoreaux}
  S.~K.~Lamoreaux,
  ``Demonstration of the Casimir force in the 0.6 to 6 micrometers range,''
  Phys.\ Rev.\ Lett.\  {\bf 78} (1997) 5.
\bibitem{Ortin}
T.~Ortin,
``Gravity and strings,''
(Cambridge University Press).
\bibitem{Spergel}
  D.~N.~Spergel {\it et al.},
  ``Wilkinson Microwave Anisotropy Probe (WMAP) three year results:
  Implications for cosmology,''
  arXiv:astro-ph/0603449.
\bibitem{Weisberg}
  J.~M.~Weisberg and J.~H.~Taylor,
  ``Relativistic Binary Pulsar B1913+16: Thirty Years of Observations and Analysis,''
  arXiv:astro-ph/0407149.
\end{thebibliography}
\end{document}